\documentclass[fleqn,twoside]{article}
\usepackage{espcrc2,isolatin1}

\usepackage{epsf}
\usepackage[figuresright]{rotating}

\newcommand{\AmS}{{\protect\the\textfont2
  A\kern-.1667em\lower.5ex\hbox{M}\kern-.125emS}}

\def\eq#1{Eq.\ (\ref{#1})}

\def\fig#1{Fig.~\ref{#1}}

\def\l{\left}
\def\r{\right}
\def\vec#1{\mathbf{#1}}

\def\msbar{{\overline{\mathrm{MS}}}}

\def\lqcd{\Lambda_{QCD}}

\newcommand{\be}{\begin{equation}}
\newcommand{\ee}{\end{equation}}
\newcommand{\bea}{\begin{eqnarray}}
\newcommand{\eea}{\end{eqnarray}}

\newcommand{\ra}{\rightarrow}

\newcommand{\cL}{\mathcal{L}}
\newcommand{\cO}{\mathcal{O}}

\def\ord#1{\cO\l(#1\r)}

\newcommand{\la}{\langle}
\renewcommand{\ra}{\rangle}

\newcommand{\gev}{\mbox{\rm GeV}}

\def\ods2{\mathcal{O}_{\Delta S=2}}
\def\zds2{Z_{\Delta S=2}}

\title{
{
\vspace{-3.3cm} \normalsize \hfill
\parbox{35mm}{\hfill CPT-2002/P.4416\\ \phantom{x}\hfill CERN-TH/2002-350
\\ \phantom{x}\hfill BUHEP-02-41} 
}\\[22mm]
$B_K$ from quenched overlap QCD\thanks{Work supported in 
part by EU HPP contract HPRN-CT-2000-00145 and US DOE grant
DE-FG02-91ER40676.  We thank Boston University for use of its
supercomputer facilities and the Gauge Connection for its gauge
configurations.}}

\author{N. Garron\address[cpt]{Centre de Physique 
Théorique$^\ddag$, Case 907, CNRS Luminy, F-13288 Marseille Cedex 9, France
\vspace{-0.2cm}},
L. Giusti$^\mathrm{a,}$
\address[cern]{Theory Division, CERN, CH-1211 Geneva 23, Switzerland
\vspace{-0.2cm}},
C. Hoelbling\addressmark[cpt],
L. Lellouch\addressmark[cpt]
\thanks{Presented by L.~Lellouch at {\it Lattice 2002}, Boston, MA.
\newline
$^\ddag$ Unité Propre de Recherche 7061.
}
and C. Rebbi\address[bu]{Department of Physics, Boston University, 590
Commonwealth Avenue, Boston MA 02215, USA
\vspace{-0.2cm}}}

\begin{document}

\begin{abstract}
We present an exploratory calculation of the standard model $\Delta
S=2$ matrix element relevant for indirect $CP$ violation in
$K\to\pi\pi$ decays. The computation is performed with overlap
fermions in the quenched approximation at $\beta=6.0$ on a $16^3\times
32$ lattice.  The resulting bare matrix element
is renormalized non-perturbatively. Our preliminary result is
$B_K^{\mathrm{NDR}}(2\;\gev)=0.61(7)$, where the error does not yet
include an estimate of systematic uncertainties.
\vspace{-0.5cm}
\end{abstract}

\maketitle

\section{Introduction}

\vspace{-0.2cm}

$K^0-\bar K^0$ mixing induces indirect CP violation in $K\to\pi\pi$
decays and is governed, in the standard model, by the 
matrix element
\be
\la\bar K^0|\cO_{\Delta S=2}(\mu)|K^0\ra=
\frac{16}{3}M_K^2\,F_K^2\times B_K(\mu)
\label{eq:rends2me}\ ,
\ee
where $\cO_{\Delta S=2}$ is the four-quark operator given below
(\eq{eq:rends2op}) and where $B_K$ parametrizes deviation from the
vacuum saturation approximation.

The benefit of using overlap fermions \cite{Neuberger:1997bg} to compute
this matrix element is their
exact $SU(N_f)_L\times SU(N_f)_R$ flavor-chiral symmetry at finite
lattice spacing
\cite{Luscher:1998pq}. This implies that the $\Delta S=2$ operator 
is simple to construct on the lattice (unlike with staggered fermions)
and does not mix with chirally dominant operators
\cite{Hasenfratz:1998jp} (as it does with
Wilson fermions). Overlap fermions further guarantee full
$\ord{a}$-improvement.  On a different note, this exploratory
computation tests the feasibility of calculating matrix elements of
four-quark operators with overlap fermions.

\vspace{-0.2cm}

\section{Simulation details}

\vspace{-0.2cm}

We use Neuberger's overlap action \cite{Neuberger:1997bg}

$$
\cL=a^4\sum_{x}\bar q\l[\l(1-\frac1{2\rho}am_q\r)
D+m_q\r]q
\ ,$$
with
$$
D=\frac{\rho}{a}\l(1+X/\sqrt{X^\dagger X}\r)\,\,\mathrm{and}
\,\, X=D_W-\rho/a
\ ,$$
where $D_W$ is the massless Wilson-Dirac operator.

The renormalized $\Delta S=2$ operator of 
\eq{eq:rends2me} is related to the bare lattice
operator through
\be
\ods2(\mu)=\zds2(a\mu,g_0)\ods2^{bare}(g_0)
\ ,
\label{eq:rends2op}
\ee
where the lattice operator with correct chiral properties is
$$
\ods2^{bare}=[\bar s\gamma_\mu(1-\gamma_5) \hat d]
[\bar s\gamma_\mu(1-\gamma_5) \hat d]
\ ,
$$
with $\hat d=\l(1-aD/2\rho\r) d$.

The computation is performed with the Wilson gauge action in the
quenched approximation at $\beta=6.0$ ($a^{-1}(r_0)\simeq2.1\,\gev$)
on a $16^3\times 32$ lattice with a statistics of 80 configurations
and with $\rho=1.4$.  The pseudoscalar mesons simulated are composed of
degenerate quarks with masses $m_q=0.04, 0.055, 0.07, 0.085, 0.1$
corresponding to a range from $\sim m_s/2\to 1.3 m_s$. Details
regarding the implementation and the inversion of the overlap operator
can be found in \cite{Giusti:2001pk}.

The three-point function used to extract the $\Delta S=2$ matrix
element and its asymptotic time behavior are
\be
\sum_{\vec{x}\vec{y}}\la J_0^L(x)\ods2^{bare}(0) J_0^L(y)\ra
\stackrel{T\gg y_0>x_0\gg 1}{\longrightarrow}
\label{eq:3ptfn}
\ee
$$
\frac{|\la K|J_0^L|0\ra|^2}{(2M_K)^2}
\la \bar K|\ods2^{bare}|K\ra
\Bigl\{\mathrm{e}^{M_K(y_0-T-x_0)}+R_K
$$
$$
\times\Bigl(\mathrm{e}^{-M_K(x_0+y_0)-\delta x_0}+
\mathrm{e}^{M_K(x_0+y_0-2T)+\delta(y_0-T)}\Bigr)\Bigr\},
$$
with $J_\mu^L{=}\bar d\gamma_\mu(1-\gamma_5)\hat s$. We allow for a
time-reversed contribution because our lattice is rather short in the
time direction and this contribution can be as large as 3-4\% for our
lightest quark mass at $x_0=T-y_0=8a$, a point which is included in
our fits. In \eq{eq:3ptfn} $\delta$ measures the finite-volume shift
in the energy of the two kaon state. This shift varies from about 2\%
to 5\% in going from our heaviest to our lightest mass and is
consistent with zero within roughly three to one standard deviations.
$R_K$ is clearly intimately related to $\la0|\ods2^{bare}|KK\ra$, but
finite volume effects can be significant and we choose not to pursue
the study of this time-reversed matrix element. The fit that we retain
for determining the forward matrix element is to times $x_0/a$ and
$y_0/a$ in the intervals $6\to 8$ and $24\to 26$ with $y_0>x_0$. To
obtain $|\la K|J_0^L|0\ra|$ and $M_K$ we compute two-point functions
of left-handed currents which we symmetrize in time and fit in the
time range $5\to 15$.

The motivation for using left-handed currents is that they eliminate
spurious zero-mode contributions which appear in finite volume and in
the quenched approximation because the overlap operator satisfies an
index theorem \cite{Hasenfratz:1998ri}, as a good Dirac operator
should. In our $J_\mu^L$ correlators, topological zero modes,
$\phi_0(x)$, contribute as
$$
\gamma_\mu(1-\gamma_5)
\frac{\phi_0(x)\phi_0^{\dagger}(y)}{m_q}\gamma_\mu(1-\gamma_5)
\ ,
$$
which vanishes because of the chirality of these modes. The use of
left-handed currents is proposed in \cite{numtech} where it is
suggested that the weak interactions of Goldstone bosons be studied in
the $\epsilon$-regime of Gasser and Leutwyler.

\vspace{-0.2cm}

\section{Chiral behavior of $B_K$}

\vspace{-0.2cm}

\begin{figure}[t]
\epsfxsize=7.cm\epsffile{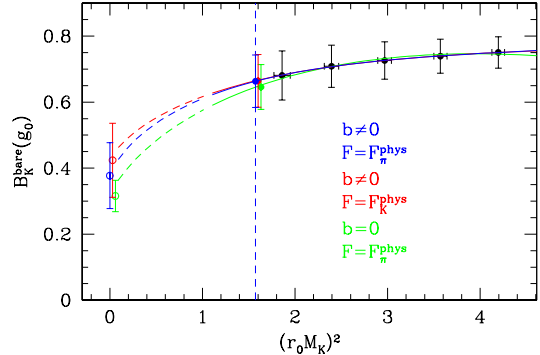}
\vspace{-1.cm}
\caption{\em Chiral behavior of $B_K$. Fits are to \eq{eq:bkchiral}, as
described in the text. The vertical dashed line indicates the physical
point.}
\vspace{-0.5cm}
\label{fig:bkchiral}
\end{figure}

We obtain $B_K^{bare}$ from
$$
B_K^{bare}=\frac38\frac{\la \bar K|\ods2^{bare}|K\ra}
{\la \bar K|\bar J_0^L|0\ra\la 0|\bar J_0^L|K\ra}
\ ,$$
where $\bar J_0^L$ is the conjugate of $J_0^L$. Quenched chiral
perturbation theory (Q$\chi$PT) predicts \cite{Sharpe:1992ft}
\be
B_K{=}B\l[1-6
\l(\frac{M}{4\pi F}\r)^2\ln
\frac{M}{\Lambda_B}
+b\l(\frac{M}{4\pi F}\r)^4\r],
\label{eq:bkchiral}\ee
where $B$, $F$ and $M$ are the leading order values of the
$B$-parameter, leptonic decay constant and pseudoscalar mass,
respectively. $\Lambda_B$ is a scale which corresponds to $\ord{p^4}$
contact terms in the effective theory and the term proportional to $b$
is added to parametrize likely higher order contributions. We fit this
functional form to our results for $B_K^{bare}$, with $M=M_K$,
$F=F_\pi^{phys}$ or $F_K^{phys}$ and $b\equiv 0$ or not. The fit
parameters are $B^{bare}$, $\Lambda_B$ and $b$ when it is not fixed to
zero.~\footnote{Note that the definition of $b$ changes with $F$.} The
results of these fits are shown in \fig{fig:bkchiral}. We find that
the mass-dependence of $B_K$ is consistent with expectations from
$\chi PT$, though results at lower masses and with better statistics
would be necessary to confirm the coefficient of the logarithm. Until
such results become available, we choose not to quote a number in the
chiral limit. At the physical point $M_K=M_K^{phys}$, however, where
the value of the $B$-parameter obtained is insensitive to the choice of
functional form, these considerations are not important.~\footnote{The
result for $B_K$ given below is obtained from the chiral fit with
$F=F_\pi^{phys}$ and $b\ne 0$.}

\vspace{-0.2cm}

\section{Non-perturbative renormalization}

\vspace{-0.2cm}

We perform all renormalizations non-perturbatively in the RI/MOM
scheme à la \cite{Martinelli:1994ty}. Thus we fix gluon configurations
to Landau gauge and compute numerically approriate, amputated forward
quark Green's functions with legs of momentum $p\equiv
\sqrt{p^2}$. Then we use the fact that with overlap fermions the
renormalization constants $Z_{J^L}=Z_V$ and $\zds2=Z_{VV+AA}$ to define
$$ 
Z_{B_K}^{\mathrm{RI}}(ap,g_0)=\frac{\Gamma_V(pa)^2} {\Gamma_{VV+AA}(pa)} 
\ ,$$ 
where $\Gamma_\cO$ is the value of the non-perturbative, amputated
Green's function of operator $\cO$ projected onto the spin-color
structure of $\cO$.

\begin{figure}[t]
\epsfxsize=7.cm\epsffile{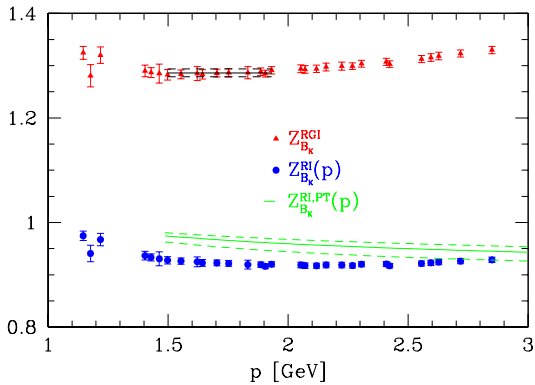}
\vspace{-1.cm}
\caption{\em Scale dependence of the renormalization constants
for $B_K$ defined in the text. The three curves for
$Z_{B_K}^{\mathrm{RI,PT}}(p)$ are obtained by allowing the scale
parameter in the coupling of \cite{Capitani:1998mq} to take the values
$\lqcd^{\msbar}/2$, $\lqcd^{\msbar}$ and $2\lqcd^{\msbar}$.}
\vspace{-0.7cm}
\label{fig:zbk}
\end{figure}

We find that in a range from approximatively 1.5 to 2 $\gev$ the
$p$-dependence of $Z_{B_K}^{\mathrm{RI}}$ matches 2-loop running 
obtained by combining the results of
\cite{Buras:1990fn} and \cite{Ciuchini:1995cd}, with $N_f=0$
and $\alpha_s$ taken from \cite{Capitani:1998mq}.  This is shown in
\fig{fig:zbk} where we plot, vs $p$, $Z_{B_K}^{\mathrm{RGI}}$ which we
define as the ratio of $Z_{B_K}^{\mathrm{RI}}$ to the 2-loop running
expression. Also shown in \fig{fig:zbk} are the non-perturbative
$Z_{B_K}^{\mathrm{RI}}(p)$ and its 1-loop counterpart
$Z_{B_K}^{\mathrm{RI,PT}}(p)$, obtained by expanding in $\alpha_s$ the
ratio of $\zds2^{\mathrm{RI,PT}}(p)$ to
$(Z_A^{\mathrm{RI,PT}})^2$. The latter are taken from
\cite{Capitani:2000da} and are properly matched.
The value of $Z_{B_K}^{\mathrm{RGI}}$ that we use is the result of
fitting our data to a constant, in the momentum range shown in the
plot. The renormalization constant in the $\msbar-\mathrm{NDR}$ scheme
is then computed through multiplying this value by the 2-loop running
expression of
\cite{Buras:1990fn} with $N_f$ and $\alpha_s$ chosen as above.

\vspace{-0.2cm}

\section{Conclusion}

\vspace{-0.2cm}

Though costly, it is entirely feasible to compute weak matrix elements
with overlap fermions, at least in the quenched approximation. The
advantages of this approach are the chiral symmetry and the
$\ord{a}$-improvement.

Our preliminary results are $Z_{B_K}^{\mathrm{RGI}}=1.29(1)$ for the 
renormalization constant and, for the $B$-parameter:
$$
\quad B_K^{\mathrm{RGI}}=0.85(10)\,,\quad B_K^{\mathrm{NDR}}(2\;\gev)=0.61(7),
$$
where the errors do not yet include an estimate of systematic
uncertainties. For the $B$-parameter, agreement with continuum-limit
world averages (see e.g. \cite{Lellouch:2000bm}) based on the
staggered result of \cite{Aoki:1997nr} is excellent.  Agreement with
the results of \cite{tom}, obtained with a different formulation of
overlap fermions, perturbative matching and larger quark masses is
good. Finally, our result is respectively a half and one standard
deviations above the results of
\cite{AliKhan:2001wr} and \cite{Blum:2001xb}, 
both obtained using domain-wall
fermions with a finite fifth dimension.


\begin{thebibliography}{9}

\bibitem{Neuberger:1997bg}
H.~Neuberger,
Phys.\ Rev.\ D {\bf 57} (1998) 5417.

\bibitem{Luscher:1998pq}
M.~Lüscher,
Phys.\ Lett.\ B {\bf 428} (1998) 342.

\bibitem{Hasenfratz:1998jp}
P.~Hasenfratz,
Nucl.\ Phys.\ B {\bf 525} (1998) 401.

\bibitem{Giusti:2001pk}
L.~Giusti {\em et al.},
Phys.\ Rev.\ D {\bf 64} (2001) 114508;
Nucl.\ Phys. {\bf 106} (PS) (2002) 739.

\bibitem{Hasenfratz:1998ri}
P.~Hasenfratz, V.~Laliena and F.~Niedermayer,
Phys.\ Lett.\ B {\bf 427} (1998) 125.

\bibitem{numtech}
L.~Giusti {\it et al.}, arXiv:hep-lat/0212012.

\bibitem{Sharpe:1992ft}
S.~R.~Sharpe,
Phys.\ Rev.\ D {\bf 46} (1992) 3146.

\bibitem{Martinelli:1994ty}
G.~Martinelli {\it et al.},
Nucl.\ Phys.\ B{\bf 445} (1995) 81.

\bibitem{Capitani:1998mq}
S.~Capitani {\it et al.}
[ALPHA Collaboration],
Nucl.\ Phys.\ B {\bf 544} (1999) 669.

\bibitem{Buras:1990fn}
A.~J.~Buras, M.~Jamin and P.~H.~Weisz,
Nucl.\ Phys.\ B {\bf 347} (1990) 491.

\bibitem{Ciuchini:1995cd}
M.~Ciuchini {\it et al.},
Z.\ Phys.\ C {\bf 68} (1995) 239.

\bibitem{Capitani:2000da}
S.~Capitani and L.~Giusti,
Phys.\ Rev.\ D {\bf 62} (2000) 114506.

\bibitem{Lellouch:2000bm}
L.~Lellouch,
Nucl.\ Phys.\ {\bf 94} (PS) (2001) 142.

\bibitem{Aoki:1997nr}
S.~Aoki {\it et al.}  [JLQCD Collaboration],
Phys.\ Rev.\ Lett.\  {\bf 80} (1998) 5271.

\bibitem{tom}
T.~DeGrand, this volume.

\bibitem{AliKhan:2001wr}
A.~Ali Khan {\it et al.}  [CP-PACS Collaboration],
Phys.\ Rev.\ D {\bf 64} (2001) 114506.

\bibitem{Blum:2001xb}
T.~Blum {\it et al.}  [RBC Collaboration],
arXiv:hep-lat/0110075.

\end{thebibliography}
\end{document}